\documentclass[aps,floatfix,nofootinbib,twocolumn,prd]{revtex4-1}

\usepackage{graphicx}
\usepackage{epic}
\usepackage{eepic}
\usepackage{latexsym}
\usepackage{amssymb}

\newcommand{\eq}[1]{(\ref{#1})}
\newcommand{\be}{\begin{equation}}
\newcommand{\ee}{\end{equation}}
\newcommand{\bea}{\begin{eqnarray}}
\newcommand{\eea}{\end{eqnarray}}

\newcommand{\hs}[1]{\hspace{#1 mm}}

\newcommand{\rr}{R^{(3)}}

\def\a{\alpha}

\def\d{\delta}

\def\e{\epsilon}

\def\f{\phi}
\def\fr{\frac}

\def\L{\Lambda}
\def\m{\mu}
\def\n{\nu}

\def\r{\rho}
\def\s{\sigma}
\def\S{\Sigma}

\def\del{\partial}

\let\bm=\bibitem
\def\nn{\nonumber}

\begin{document}

\title{Remarks on inhomogeneous anisotropic cosmology}

\author{Ali Kaya}
\email{ali.kaya@boun.edu.tr}
\affiliation{ Bo\~{g}azi\c{c}i University, Department of Physics, \\ 34342,
Bebek, Istanbul, Turkey }

\begin{abstract}

Recently a new no-global-recollapse argument was given for some inhomogeneous and anisotropic cosmologies that utilizes surface deformation by the mean curvature flow. In this paper we discuss important  properties of the mean curvature flow of spacelike surfaces in Lorentzian manifolds. We show that singularities may form during cosmic evolution and the theorems forbidding the global recollapse lose their validity. The time evolution of the spatial scalar curvature that may kinematically prevent the recollapse is determined in normal coordinates, which shows the impact of inhomogeneities explicitly.  Our analysis indicates a caveat in numerical solutions that give rise to inflation. 

\end{abstract}

\maketitle

\section{Introduction}

Inflation is the leading early universe paradigm; it offers a consistent cosmic evolution and inherently yields an almost scale invariant cosmological perturbations as the seeds of the structure. However, the basic mechanism driving inflation and, correspondingly, the naturalness of its beginning are still questionable (see \cite{r1} for an early review and \cite{rb} for a recent review). There are different approaches to the initial condition problem of inflation. For example, one may introduce a probability measure in the phase space and try to determine the possibility of inflation from the set of  allowed initial conditions. In the minisuperspace approximation, a viable canonical measure with suitable properties has been proposed in \cite{p1} but even in this simplified setting the result turned out to be inconclusive as the measure of the region in the phase space giving inflation diverges \cite{p2}. In \cite{p3}, a physically viable cutoff has been proposed to obtain a finite probability for inflation, but as noted in \cite{p4} the region restricted by the cutoff deforms nontrivially under the Hamiltonian evolution, which weakens the naturalness of the idea (see also \cite{p5} for other issues). 

According to the general lore, to realize inflation it is enough to have a ``small" homogeneous and isotropic region dominated by an (effective) cosmological constant. Indeed, inflation seems inevitable in the presence of a cosmological constant since in an expanding universe it eventually dominates other sources.  Supportively, it is known that in the presence of a cosmological constant all initially expanding flat and negative spatial curvature homogeneous but anisotropic Bianchi cosmologies exponentially asymptote to de Sitter spacetime \cite{w1}. However, this result has no simple generalization to other cases. For spatially closed cosmologies, there are theorems characterizing the final state of the universe  \cite{cu2}; nevertheless, there is no simple answer even for the closed Friedmann spacetime (see e.g. \cite{cu1, cu3}). The matter gets more complicated since the inflationary dynamics is supposedly governed by an effective cosmological constant that must disappear in time and it is known that some simple scalar field inflationary models contain the so-called sudden singularities \cite{s1,s2,s3,s4}. Besides, inflation is thought to be occurring in an ambient space and thus surface effects may destroy the expansion since the interface contracts towards the inflating region (note that while the inflating region has negative pressure the outside has presumably positive pressure) \cite{em1}.  There are also obstructions in embedding an initially subhorizon inflating region into a decelerating Friedmann universe imposed by the null geodesic propagation, which shows that in such a setup inflation also needs large scale homogeneity \cite{em2}. As discussed in \cite{em3}, the boundary effects may even spoil the scale freeness of the cosmological perturbations. 

In an interesting recent paper \cite{k1}, it is argued that if the weak energy condition is satisfied and each time slice has a region with negative spatial scalar curvature, the most general initially expanding inhomogeneous and anisotropic cosmology cannot  recollapse everywhere.  The argument is based on the Gauss-Codazzi equation, which in this context becomes the Hamiltonian constraint of general relativity, and the mean curvature flow of the time slices. One may see that the mean curvature flow of any given time slice either yields an extremal surface with zero mean curvature everywhere or strictly increases the volume of the surface. When the weak energy condition is satisfied and the spatial scalar curvature is negative somewhere, the first outcome is forbidden by the Gauss-Codazzi equation. Hence, the mean curvature flow gives spatial sections with increasing volume and a total recollapse is forbidden. In \cite{k1}, a lower bound for the expansion rate is also found. 

Curiously, the argument given in \cite{k1} only uses the weak energy condition and one of the constraint equations of general relativity. Namely, it seems that the result does not depend on the details of the dynamical evolution. One may compare the situation with the singularity theorems that either suppose the strong energy condition controlling the dynamics; or the null energy condition, which restricts the constraints, and an additional dynamical assumption, i.e. the existence of a trapped surface. This curiosity is the main motivation of the present work. 

\section{Mean curvature flow}

For a given spacelike surface embedded in a spacetime, the mean curvature is given by $K=\nabla_\m n^\m$, where $n^\m$ is the future directed unit normal vector to the surface and $\nabla_\m$ is the covariant derivative of the spacetime metric. The mean curvature flow is defined by deforming the surface along the integral lines of the vector field $Kn^\m$, i.e. in local $x^\m$ coordinates, it is given by 
\be\label{mcf}
\fr{dx^\m}{ds}=Kn^\m,
\ee
where $s$ parametrizes the flow  (not to be confused with the geodesic parameter).  As pointed out in \cite{k1}, the deformed surface has strictly larger volume. It is not difficult to understand this result intuitively as the flow deforms a local region forwards in time if it is expanding ($K>0$) or backwards in time if it is contracting ($K<0$). The point on the surface with $K=0$ is an {\it instantaneous} fixed point of the flow but as the nearby region potentially bends, $K$ may no longer remain zero. 

In Euclidean signature the mean curvature flow is defined to reduce the volume and may become singular in finite flow time for nonconvex initial data \cite{mcf0}. In Lorentzian  signature, the flow is much more regular since it increases the volume. In a cosmological spacetime with compact spatial sections, the finite time existence proof is given in \cite{mr2} provided the timelike convergence condition 
\be\label{se}
R_{\m\n}k^\m k^\n\geq0
\ee 
is satisfied for all timelike vectors $k^\m$, where $R_{\m\n}$ is the Ricci tensor (see section 4 of \cite{mr2}). Assuming Einstein's equations 
\be\label{ee}
G_{\m\n}=T_{\m\n},
\ee
\eq{se} becomes the strong energy condition for matter. In \cite{mr3},  \eq{se} is relaxed so that in some cases the weak energy condition can be enough for the regularity. The flow can be continued as long as the deformed surface remains smooth and compact. Supposing the existence of two barrier surfaces, which are located to the past and to the future of the initial surface with positive $K>0$ and negative $K<0$  mean curvatures everywhere, the flow \eq{mcf} exists for all times and converges uniformly to a spacelike surface with  vanishing mean curvature $K=0$ everywhere. The existence of such a surface has originally been proved by a different method that does not require any condition of the sort \eq{se}; see theorem 6.1 of \cite{mr1}.  

\begin{figure}
\centerline{\includegraphics[width=8.5cm]{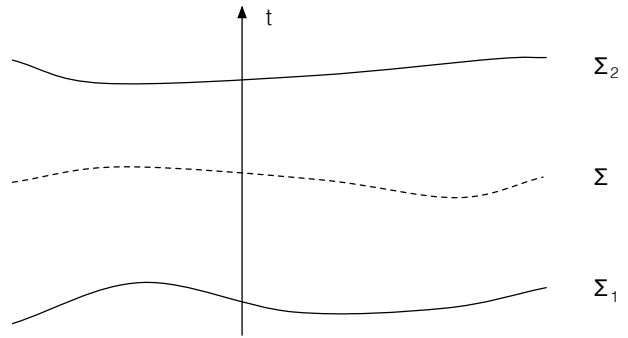}}
\caption{A schematic explanation of the main theorems proved in \cite{mr2,mr3,mr1}: In a smooth spacetime, assuming that two barrier timeslices $\Sigma_1$ and $\Sigma_2$ exist with respective mean curvatures $K_1(\s^i)$ and $K_2(\s^i)$ satisfying $K_2<K_1$,  then for any given function $K(\s^i)$ strictly obeying $K_2<K<K_1$ there exists a timeslice $\S$ whose mean curvature is equal to $K(\s^i)$ .}
\label{fig2}
\end{figure}

The  proofs of the theorems given in \cite{mr2,mr3,mr1}, which are based on the theory of nonlinear (parabolic) differential equations, are rather technical and not easy to follow. To illustrate the main result in a simple example, let us consider a smooth cosmological spacetime with the metric $ds^2=-dt^2+a(t)^2 dx^idx^i$. The mean curvature of a constant time hypersurface is given by $K=3\dot{a}/a=3H$. If one assumes the existence of two barrier hypersurfaces with times $t_1$ and $t_2$ obeying $H_2<H_1$, then the theorems show the existence of a timeslice with the mean curvature $K=3H_*$, where $H_2<H_*<H_1$. In this example, this result  obviously follows from the continuity of the Hubble parameter $H$. For nonconstant mean curvatures, the theorems are illustrated in Fig. \ref{fig2}. 

Locally, the mean curvature flow is given by the functions $x^\m(s,\s^i)$, where $\s^i$ are the intrinsic coordinates of the surface. The induced metric and the extrinsic curvature are defined as
\bea
&&h_{ij}=\del_ix^\m\del_jx^\n g_{\m\n},\nn\\
&&K_{ij}=\del_ix^\m\del_jx^\n\nabla_\m n_\n,\label{imec}
\eea
where $\del_i=\del/\del\s^i$ and $g_{\m\n}$ is the spacetime metric. The extrinsic curvature can be decomposed as
\be
K_{ij}=\fr13 h_{ij}K+\s_{ij}
\ee
where $K$ is the mean curvature and $\s_{ij}$ is the shear. A straightforward calculation\footnote{For example, in getting the first equation in \eq{fe} one may take the $s$-derivative of  the first equation in \eq{imec}. One then observes $d/ds (\del_i x^\m)=\del_i (dx^\m/ds)=\del_i (Kn^\m)$, where the flow equation \eq{mcf} is used. As $n_\m\del_ix^\m=0$, i.e. $n^\m$  is perpendicular to $\del_i x^\m$ which is tangent to surface, the $\del_i$ derivative can be moved to the other terms, i.e. $\del_i (Kn^\m) \del_j x^\n g_{\m\n}=-Kn^\m\del_i(\del_j g_{\m\n})$. Collecting all terms, one arrives at the first equation in \eq{fe}. Note that there is no assumption about $n^\m$ being geodesic.} yields
\bea
&&\hs{-10}\fr{dh_{ij}}{ds}=2KK_{ij},\nn\\
&&\hs{-10}\fr{dn_\m}{ds} \del_ix^\m =\del_iK. \label{fe}\\
&&\hs{-10}\fr{dK}{ds}=D_iD^iK-\fr13 K^3-K(\s^{ij}\s_{ij}+R_{\m\n}n^\m n^\n),\nn
\eea
where $D_i$ is the covariant derivative of $h_{ij}$. The bitensor $\del_ix^\m$ projects vectors of spacetime into the tangent space of the surface. From \eq{fe}, one sees that the variation of the normal vector is proportional to $\del_iK$.  In the evolution equation of $K$, the first term gives a diffusion  effect  that smooths out $K$. The $K^3$ term forces $|K|$ to decrease in the form of a power law in the flow parameter. The last term explicitly shows the importance of the timelike convergence condition \eq{se}, which  may otherwise yield an exponential instability for $K$. 

From the first equation in \eq{fe}, one sees that
\be\label{vi}
\fr{d\ln(\sqrt{h})}{ds}=K^2,
\ee
where $h=\textrm{det}\,h_{ij}$. Consequently the local volume of the surface increases with the flow. This last equation also shows that {\it in a smooth evolution} the mean curvature flow preserves the causal structure of the initially spacelike surface, because otherwise the timelike normal vector must first become null and the induced metric of a null surface degenerates so that $h=0$, which is impossible since $h$ increases with the flow.   

\begin{figure}
\centerline{\includegraphics[width=5cm]{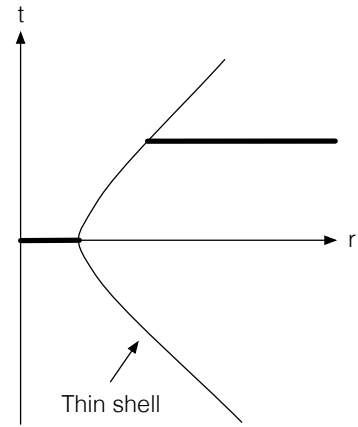}}
\caption{The motion of an infinitesimally thin spherical shell that separates an empty flat space (left of the hyperbola) placed inside a de Sitter space (right of the hyperbola)  as seen by an observer in the flat space. The thick black lines give the mean curvature flow of the $t=0$ surface.}
\label{fig1}
\end{figure}

It is important to emphasize that the above results assume a {\it smooth} evolution that essentially requires a {\it smooth} background. For example, one may find {\it singular} spacetimes where the mean curvature flow fails to preserve the causal structure. Consider the analytic solution given in \cite{em3} that describes an infinitesimally thin spherical shell with empty flat interior evolving in a de Sitter spacetime. Physically, the solution describes a true vacuum bubble immersed in a false vacuum region and the dynamics can be analyzed by the junction conditions of \cite{j}.  In the flat space coordinates, the trajectory of the shell radius corresponds to the world line of a particle with constant acceleration, which is explicitly given by 
\be
r=\fr{1}{\a}\cosh(\a t),\hs{5}t=\fr{1}{\a}\sinh(\a t),
\ee
where $\a$ is fixed by the surface tension and the Hubble parameter of the de Sitter space \cite{em3}. The region to the left of the hyperbola is flat and the right  expands exponentially (in fact, for the right-hand side a new coordinate system must be introduced; see \cite{em3}). In this spacetime, think about the mean curvature deformation of the $t=0$ surface, which has zero mean curvature ($K=0$) for $r<1/\a$ and constant positive mean curvature ($K>0$) when $r>1/\a$. Since $K$ is discontinuous across the shell, its derivative gives  a Dirac delta function and the (nonlinear) flow equations \eq{fe} become ill defined (for instance, they involve the powers of the Dirac delta function). In that case, the only viable way to define the flow is to use the main defining equation \eq{mcf} independently in both regions, in the spirit of \cite{j}.  Now, in a FRW background with the metric $ds^2=-dt^2+a(t)^2dx^idx^i$, the normal vector to a timeslice is $n^\m=\d^\m_t$ and the mean curvature becomes $K(t)=3\dot{a}/a$. Therefore, \eq{mcf} reduces to $dt/ds=K(t(s))$, which can be solved as $t=t(s)$, i.e. the mean curvature flow deforms an initial timeslice to another one (forward or backward in time depending on the sign of $K$). The {\it singular shell} can be viewed to cut the surface and while the interior is fixed under the flow (since $K=0$ entirely), the exterior is pushed arbitrarily forward in time changing the causal structure of the initial surface; see Fig. \ref{fig1}.  

For any given spacelike surface, one can introduce (Gaussian) normal coordinates $(t,x^i)$ so that the metric becomes 
\be\label{met}
ds^2=-dt^2+h_{ij}(t,x)dx^i dx^j,
\ee
and the surface is described by the equation $t=t_0$.  The extrinsic and the mean curvatures of constant time slices can be found as
\be\label{exc}
K_{ij}=\fr12\dot{h}_{ij},\hs{5}K=\fr12h^{ij}\dot{h}_{ij},
\ee
where the dot denotes the partial time derivative. From \eq{mcf}, the first iteration of the mean curvature flow of $t=t_0$ surface gives  
\be
t\simeq t_0+K(t_0,x)\d s ,
\ee
where $\d s$ is chosen as small as possible. The unnormalized normal vector of the new surface is found as $N_0=-1$ and $N_i=\del_iK\d s$ whose length is given by 
\be\label{N}
N^\m N_\m=-1+h^{ij}\del_iK\del_jK\d s^2.
\ee
One sees that the normal is tilted more towards the light cone when $\del_iK$ is large. Similarly, the mean curvature flow maps the two infinitesimally nearby points on the initial surface $(t_0,x^i)$ and $(t_0,x^i+\e^i)$ to
\bea
&&\hs{-10}(t_0,x^i)\to(t_0+K(t_0,x^i)\d s,x^i),\nn\\
&&\hs{-10}(t_0,x^i+\e^i)\to(t_0+K(t_0,x^i+\e^i)\d s,x^i+\e^i).\nn
\eea
The vector connecting these two points, which becomes tangent to the deformed surface as $\e^i\to0$, can be found as 
\be
k^\m\simeq(\e^i\del_iK\d s,\e^i).
\ee
From the norm $k^\m k_\m\simeq h_{ij}\e^i \e^j-(\e^i\del_iK)^2\d s^2$, one again sees that the surface is tilted more with respect to the reference spacelike slice with increasing 
\be
\fr{1}{\sqrt{h_{ij}\e^i\e^j}}|\e^i\del_iK|,
\ee
which is the directional derivative of $K$.

\section{Spatial scalar curvature}

The argument of \cite{k1} is based on the existence of a region with negative spatial scalar curvature in any given time slice. This can be ensured topologically if the spacetime is globally $\mathbb{R}\times \S_3$ and the three-dimensional space $\S_3$ is compact, oriented and topologically in the class of open or flat manifolds as considered in \cite{k1}. The region having negative spatial curvature is by no means fixed in $\S_3$ and different time slices may have different parts having that property. Namely, the negative spatial curvature region wanders in $\S_3$.

Given a spacetime that has the form  $\mathbb{R}\times \S_3$, one can introduce normal coordinates $x^\m=(t,x^i)$ adapted to $\S_3$ so that  the metric locally becomes
\be\label{m}
ds^2=-dt^2+h_{ij}(t,x)dx^i dx^j.
\ee
Using that the extrinsic curvature of a constant time surface is given by \eq{exc}, the Einstein's equations \eq{ee} yield   
\bea
&&\rr+\fr23 K^2-\s^{ij}\s_{ij}=2\rho\label{hc},\\
&&D^j\s_{ji}-\fr23 \del_iK=T_{0i}\label{mc},
\eea
which are the Hamiltonian and momentum constraints respectively, and 
\bea
&&\hs{-5}\dot{K}_{ij}-\dot{K}h_{ij}=T_{ij}-G^{(3)}_{ij}-KK_{ij}+2K_{ik}K^{k}{}_{j}\nn\\
&&+\fr12 h_{ij}\left( K^2+K^{kl}K_{kl}\right),\label{de}
\eea
which is the time evolution equation.  Here, all index manipulations are carried out by $h_{ij}$; $\rr$ and $G^{(3)}_{ij}$ are the Ricci scalar and Einstein tensors of $h_{ij}$, $D_i$ is the covariant derivative of $h_{ij}$, $T_{\m\n}$ is the total energy momentum tensor possibly including a cosmological constant and $\r=T_{00}$ is the total energy density. In these coordinates, the conservation equation $\nabla_\n T^{\m\n}=0$ decomposes into 
\bea
&&\dot{\r}+D_iT^{0i}+K\r+K_{ij}T^{ij}=0.\nn\\
&&\dot{T}_{0i}+KT_{0i}-D^jT_{ji}=0.\label{encon}
\eea
Once the initial conditions for the metric and the matter fields satisfying the constraints \eq{hc} and \eq{mc} are imposed, \eq{de} can be used to determine the time evolution. 

From \eq{hc}, one sees that as long as the weak energy condition is obeyed by matter so that $\rho \geq0$, $K$ cannot vanish when $\rr<0$. Thus, an expanding region with $\rr<0$ cannot  start contracting later. However, locally $\rr$ may increase in time and become positive, which may turn the local expansion into a contraction. Of course, due to the specified topology of $\S_3$, $\rr$ cannot be  positive everywhere so some other region in the slice must have negative $\rr$. If the evolution continues {\it smoothly}, the global contraction with $K<0$ is forbidden since the theorem 6.1 of \cite{mr1} requires the existence of a $K=0$ timeslice, which is impossible by \eq{hc} since $\rr<0$ somewhere. However, {\it singularities} may form in regions that start contracting, while some other regions still expand. Then, the theorem of \cite{mr1} loses its validity (since it assumes smoothness to a certain order), and unless it is generalized to deal with these singularities\footnote{According to the cosmic censorship hypothesis,  singularities are hidden within event horizons. Still, the theorem of \cite{mr1} loses its validity since the spacetime can no longer be globally hyperbolic, at least when a Kerr black hole forms.} there is no obvious obstruction for an initially expanding spacetime to later develop singularities and become totally contracting. 

As noted in \cite{k1}, the Hamiltonian constraint \eq{hc} gives the bound $|K|\geq\sqrt{3\r}$ that restricts the expansion or the contraction rate of a negative $\rr$ region. Therefore, in topologically flat or open cosmologies there must always be a region obeying this bound. In an initially expanding universe, if it had been the case that $\dot{K}<0$, then it would be possible to deform $K<0$ regions of a given slice backwards in time to obtain a new slice with negative $K$ arbitrarily close to zero. Then, the only way to satisfy the bound $|K|\geq\sqrt{3\r}$ would be to assume the existence of an expanding region with $K\geq\sqrt{3\r}$.  But, from \eq{de} one may find that 
\be\label{dk}
\dot{K}=\fr12\rr-\fr32\r-\fr12h^{ij}T_{ij}-\fr32\s^{ij}\s_{ij}.
\ee
If the null energy condition $\r+P\geq0$ is satisfied, all but the first term in the right-hand side are strictly negative. Nevertheless, a positive $\rr$ may still yield $\dot{K}\geq0$ and in that case {\it uniformly} deforming $K<0$ regions backwards in time  give a surface with larger $K$.  

In the presence of a cosmological constant $\L$, the energy density $\rho$ has a minimum. In that case, one has the following argument \cite{k1}: At early times in a globally expanding universe there must exist a time slice with $K>\sqrt{3\L}$. If at late times one finds a time slice with $K<\sqrt{3\L}$, then \cite{mr1} shows the existence of a time slice with $K=\sqrt{3\L}$, which is forbidden by \eq{hc} when $\rr<0$ (the early and the late surfaces with $K>\sqrt{3\L}$ and $K<\sqrt{3\L}$ are barrier surfaces in the theorem 6.1 of \cite{mr1}). This naively shows that if there is a cosmological constant then each time slice has a region expanding as fast as the corresponding de Sitter space \cite{k1}. However, again, the possible emergence of singularities invalidates the theorem of \cite{mr1}, and consequently this conclusion is not necessarily correct. 

As pointed out in the Introduction, a curious feature of  the no-global-recollapse argument of \cite{k1} is that it only uses the weak energy condition and one of the constraint equations of general relativity. The above results show that the dynamical evolution equations cannot be ignored in such an analysis. The theorems of \cite{mr2,mr3,mr1} assume the existence of a smooth spacetime metric, which is not necessarily respected by the generic dynamical evolution. In essence, the smoothness requirement is an extra  dynamical assumption that leads to the no-global-recollapse result. 

Since the spatial scalar curvature plays the crucial role in this discussion, it is useful to determine its time evolution. From the time derivative of \eq{hc}, we obtain a surprisingly simple result
\be\label{rt}
\dot{R}^{(3)}=-\fr23K\rr-2\s^{ij}R^{(3)}_{ij}-2D_iT^{0i}.
\ee
The first term dictates the scaling of $\rr$ under time evolution; i.e. $|\rr|$ decreases or increases under expansion or  contraction, respectively (in a Friedmann universe it gives the standard $k/a^2$ behavior of the constant spatial curvature). Evidently, this term cannot change the sign of $\rr$. Since $\s_{ij}$ is traceless, the second term is independent of $\rr$, and thus the last two terms in \eq{rt} can be viewed as ``external sources" for $\dot{R}^{(3)}$. 

From \eq{encon} one sees that the last term in \eq{rt} gives the local energy inflow. As discussed in \cite{w1} one expects  a positive inflow towards an inflating region, which enforces $\rr$ to increase.  Similarly, the off-diagonal components of  $R^{(3)}_{ij}$ are related to matter stress, and together with the shear $\s_{ij}$, they also affect the time rate of change of $\rr$. While the first effect can be ascribed to inhomogeneity, the second can be viewed as the impact of anisotropy. In an inhomogeneous anisotropic universe, both of these terms can be large and potentially give large deviations of $\rr$ in time. 

The above analysis raises the following concern about the numerical solutions yielding inflation. We saw that as long as $\rr<0$, an initially expanding region cannot recollapse. When the region expands, the cosmological constant dominates all other sources and therefore inflation has a good chance to start provided $\rr$ stays negative. On the other hand, $\rr$ can only change sign if the last two terms in \eq{rt} are effective. Therefore, by incidentally choosing initial conditions so that the last two terms in \eq{rt} start out small, one artificially increases the possibility of inflation. This is the case\footnote{In \cite{num},  the code given in \cite{num2} is used to solve the initial constraint equations, which assumes a conformally flat initial spatial metric. Moreover, \cite{num2} uses ``superposed free data" as they are interested in initially isolated compact objects. It is not clear to us whether these assumptions are suitable for an early cosmological setup.} in the recent numerical work \cite{num}, which initially sets $\s_{ij}=0$ and $T_{0i}=\dot{\f}\del_i\f=0$.

Another related concern is that the constraint equations do not fix the initial values of $R^{(3)}_{ij}$ and $\s_{ij}$ uniquely. Assuming that the initial conditions of matter are first specified, which is usually the case in numerical analysis, the constraint equations are solved for the initial metric data \cite{num0}, but there still exists some remaining freedom. This is evident in normal coordinates where the constraint equations \eq{hc} and \eq{mc} only determine $\rr$. Of course, the choice \eq{m} does not fix all diffeomorphism invariance, yet the remaining freedom is larger than the residual gauge symmetry: After solving all constraints and fixing all diffeomorphism invariance, the metric  still has two independent degrees of freedom, which must be specified by imposing physically viable boundary conditions. The danger here is that one may incidentally choose initial data without paying attention to this freedom so that $R^{(3)}_{ij}$ and $\s_{ij}$ become small, which may undermine the evolution of $\rr$ significantly. This is the case in \cite{num1}, where the initial metric is chosen to be flat $h_{ij}=\d_{ij}$ and the initial shear is set to be zero $\s_{ij}=0$. Clearly, these choices are questionable in the presence of inhomogeneities and anisotropies. 

The above comments do not  necessarily dispute the results obtained in the numerical simulations.  Specifically, in \cite{num} the initially chosen large inhomogeneities and anisotropies presumably give large contributions to \eq{rt} later on. But still, the expansion and the matter flow rates are, in general, different than each other, so this can only be verified by simulations;  we think it would be interesting to check numerically the impact of the last two terms in \eq{rt} at the beginning of inflation. As suggested in \cite{k1}, one may attemp to view their no-global-recollapse argument as an analytic explanation of the numerical results of \cite{num}. Yet, we see that the argument of \cite{k1} is only valid under certain technical conditions that might be violated during cosmological evolution. Moreover, \cite{k1}  uses a {\it topological restriction} for timeslices so that the spatial scalar curvature {\it cannot} be positive everywhere. How such a topological condition can be implemented and affect a numerical simulation is not obvious.

\section{Conclusions}

Determining the naturalness of inflation is a crucial open problem and we think that the arguments of \cite{k1} give significant guidance towards a solution. The spatial scalar curvature seems to play the key role in this problem. As the classification of the three-dimensional closed manifolds is known, it would be interesting to generalize the story of constant curvature flat, open and closed cosmologies to other possibilities. The theorems of \cite{mr2,mr3,mr1}, which show the existence of time slices with prescribed mean curvatures, can be used as powerful tools in cosmology to infer interesting results. However,  these theorems assume certain smoothness properties that may be absent in realistic scenarios. 

Our analysis also indicates a possible delicacy in selecting the initial conditions in numerical solutions; some simplifying assumptions might  generate hidden effects favoring a dynamical behavior. Since solving the constraints and fixing the coordinates do not specify the initial spatial metric uniquely, it would be interesting to investigate how this arbitrariness can be fixed. Given the matter distribution, this usually requires the imposition of physically viable boundary conditions like the fall-off conditions encountered in the asymptotically flat spacetimes. Another interesting problem is to find a way of characterizing inhomogeneities and anisotropies. For instance, from \eq{rt} one is able to pin down and interpret the terms affecting the time evolution of $\rr$. It might be useful to generalize this analysis to other geometric quantities  to understand how inhomogeneities and anisotropies affect their evolution. 

\section*{Acknowledgments} 

I would like to thank M. Kleban  and J. Barrow for useful comments and discussions.

\end{document}